\documentclass[preprint,showpacs,amsmath,prb]{revtex4} 
\usepackage{graphicx}

\begin{document}

\title{New type of lenses based upon left-handed materials}
\author{A. L. Pokrovsky}
\affiliation{University of Utah, Salt Lake City UT, 84112 USA}
\author{A. L. Efros}
\affiliation{University of Utah, Salt Lake City UT, 84112 USA}

\begin{abstract}
New type of lenses which are a slab of a left-handed material
embedded into a regular material is proposed. 
These two materials should have equal refractive indices.
Lenses with a focal length larger than the width of the 
slab can be constructed.
These lenses should be easier to make than the well known 
Veselago lens, because the materials of the Veselago lens 
should obey an additional matching condition.  
Lenses of new type have multiple foci and might be useful for the 3D imaging.
\end{abstract}
%\ocis{000.2690, 080.3620, 120.5700, 120.5710}

\maketitle

In his seminal work Veselago \cite{ve} has introduced the concept
of left-handed materials (LHM's).
In a simplest case the LHM's are materials 
with simultaneously negative
electric permittivity $\epsilon$ and 
magnetic permeability $\mu$ in some frequency range. 
It is easy to show that in LHM the vectors ${\bf k}, {\bf E}, {\bf H}$
form a left-handed set, while in usual materials ($\epsilon > 0$, $\mu > 0$)
they form a right-handed set.
If the imaginary parts of $\epsilon$ and $\mu$ are small,
the electromagnetic waves (EMW's) propagate in the LHM but they have some 
unusual properties. All these properties originate from the fact that
in the isotropic LHM the Poynting vector 
${\bf S} = {\bf E} \times {\bf H}$ is anti-parallel to the 
wave vector ${\bf k}$.

\begin{figure}
\includegraphics[width=8.6cm]{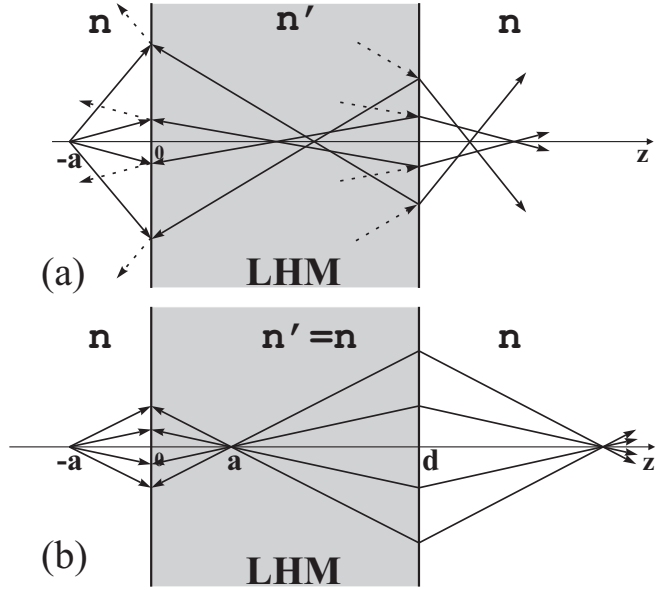}
\caption{Reflection and refraction of light outgoing from a 
point source at $z=-a$ and passing through the slab of the LHM at $0<z<d$.
Refraction of light is described by the anomalous Snell's law.
The arrows represent the direction of the wave vector.
The reflected waves are shown by dashed lines near each interface only.
The slab is surrounded by the usual RHM. (a) $n' > n$. 
(b) The Veselago lens ($n' = n$). The reflected waves are absent, 
all rays pass through two foci.}
\label{fig1}
\end{figure}

Consider a  propagation of the EMW from a point source located 
at the point
$z = -a$ through an infinite slab of the 
LHM with the thickness $d$ and a 
usual right-handed material (RHM) at $z<0$ and $z>d$ (Fig.\ref{fig1}).
We assume below that refractive indices of the LHM and the RHM are the same.
This assumption excludes the total internal reflection. Then
transmission coefficient of the slab is non-zero and
it is obvious that $S_z > 0$ everywhere at $z > -a$
because the energy propagates from its source.
Directions of the vector ${\bf k}$ for different rays are shown by arrows. 
They  should be  chosen in such a way that at both
 interfaces tangential components of vector ${\bf k}$
for incident, reflected and refracted waves are the same. Another condition is 
that the component $k_z$ should be parallel to $S_z$ in the RHM 
and anti-parallel in the LHM.
%Then in the LHM $k_z$ is negative.
It follows that the Snell's law for the RHM-LHM interfaces
has an anomalous form:
$\sin i / \sin r = -n'/n$, where $i$ and $r$ are the angles of incidence and
refraction respectively, $n' = \sqrt{|\epsilon '| |\mu '|/\epsilon_0 \mu_0}$ 
and $n = \sqrt{\epsilon \mu/\epsilon_0 \mu_0}$ are positive refractive indices
for LHM and RHM respectively. 
The angles of reflection are equal to the angles of incidence.
Refractive index in the LHM's is often defined 
as negative.\cite{nv,va,p2,sm3}
It has been shown recently that introduction 
of the negative refractive index for the LHM's 
is unnecessary and even misleading.\cite{nnotneg}

The device shown in Fig.\ref{fig1}(b) is a unique 
optical lens proposed by Veselago.\cite{ve} 
In this lens  $\epsilon = -\epsilon '$ and $\mu = -\mu '$, then 
$n' = n$ and $i = -r$.
It is easy to show that in this case the reflected wave is completely absent.
Since all rays going right from the source have $i=-r$,
all of them have foci at points $z=a$ and $z=2d-a$ as shown 
in Fig.\ref{fig1}(b).

Recently a method of fabricating of the left-handed metamaterials on the
basis of metallic photonic crystals has been found and 
the San Diego group has reported the first observation
of the anomalous transmission \cite{sm} 
and even the anomalous Snell's law.\cite{sm3}
Both observations have been interpreted as the result
of negative $\epsilon$ and $\mu$.
The speculations about the nature of 
negative $\epsilon$ and $\mu$ in the proposed metamaterials
are still controversial\cite{sm3,p2,mar,loww}, 
but the very existence of the LHM seems to be demonstrated.  

Pendry has proposed that the Veselago lens (VL) is a perfect lens
in a sense that the width of its foci does not have
usual wave length limitation. 
This is still a controversial point.\cite{p2,nv,com1,com2,haldane,lens} 
Our paper
is based upon geometrical optics only so that this unresolved 
question does not appear.

\begin{figure}
\includegraphics[width=8.6cm]{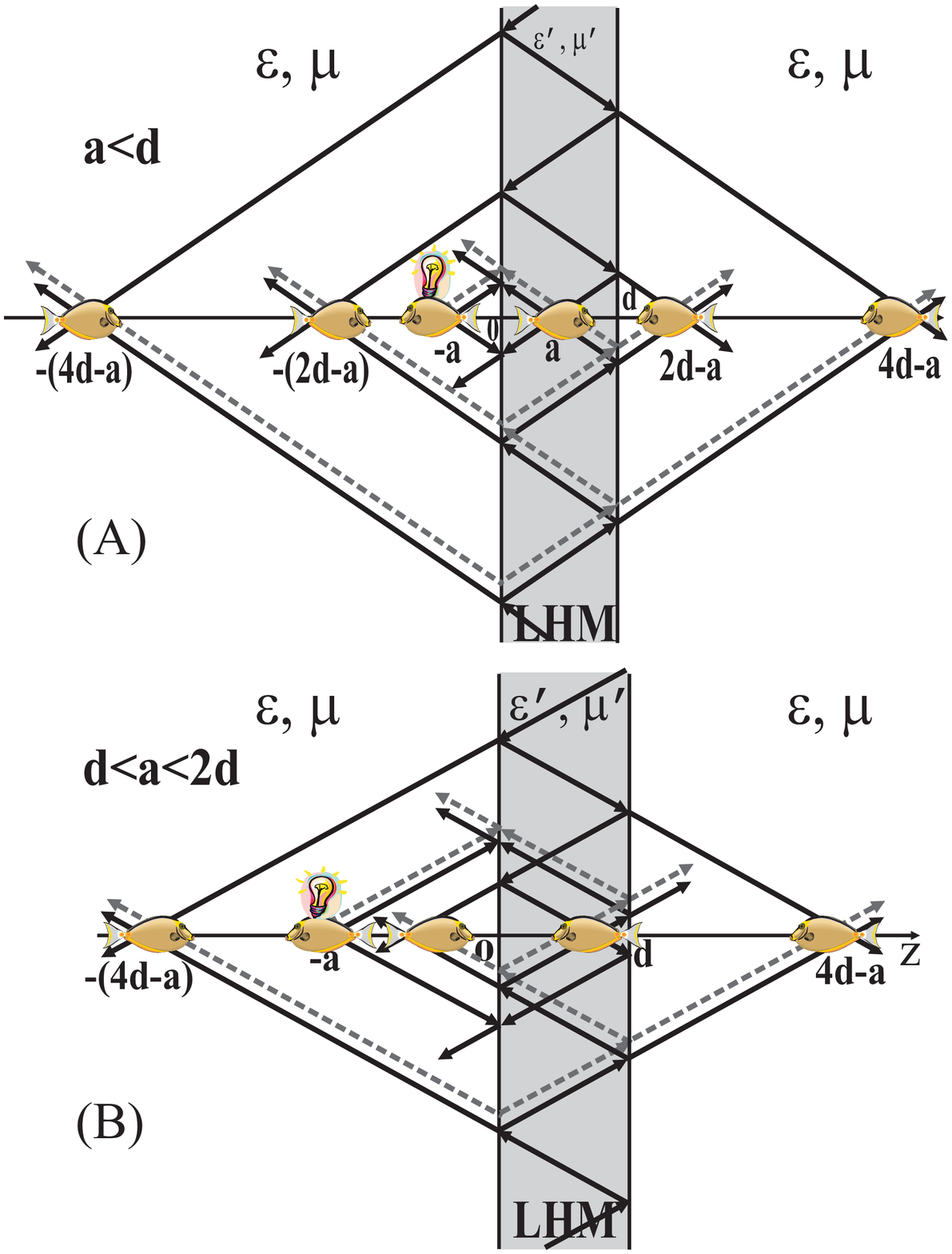} \vspace{0.5cm}
\caption{Multiple 3D images (``fishes'') of the object marked
by the bulb. The active element of the lens is a slab made of the LHM with 
$\epsilon' \mu' = \epsilon \mu$, but $\epsilon' \neq -\epsilon$ and $\mu' \neq -\mu$.
The arrows show the directions of the wave vectors, which are opposite
to the Poynting vector inside the slab. 
(A) $a<d$; (B) $d<a<2d$. }
\label{fig2}
\end{figure}

In this paper we propose and analyze another type of lenses, 
which has geometrical construction similar to the VL.
In these lenses $\epsilon' \mu' = \epsilon \mu$ like in the VL,
but $\epsilon' \neq -\epsilon$, $\mu' \neq -\mu$.
Since $n'= n$ the absolute value of the angles of incidence, reflection
and refraction is the same and this system may also work as a lens
(see Fig. \ref{fig2}).
The most important difference between the new type of lenses
and the VL is the presence of reflected waves. 
As a consequence, the new lenses produce a periodic array of 3D 
images of different intensity.
The lenses should be a lot easier to manufacture than the VL, because 
the materials for them should obey one condition only 
($\epsilon \mu = \epsilon' \mu'$) instead of
two in the case of the VL ($\epsilon = -\epsilon'$, $\mu = -\mu'$).
Another advantage is that the new lenses can have a focal length greater than 
the width of the slab $d$ and produce images
of objects located at distances larger than $d$ (Fig. \ref{fig2}(B)).
In contrast, VL produces real images of the objects 
located at a distance closer than $d$ from the lens only.
%Another difference is that VL produces only one image of the 
%object outside the lens, while new lenses
%create an array of foci (Fig. \ref{fig2}).

One can see that both the VL and the new lenses
are {\it absolute instruments} because they image 
stigmatically three-dimensional domains and the 
optical length of any curve in the object space is equal to 
the optical length of their images.\cite{born}
Note, that the above properties of the absolute instrument
should be valid in the limit of geometrical optics only.
For the VL this 3D domain is limited by the condition $-d \le z \le 0$, 
while for the new lenses it is the whole half-space $z<0$.
Since the LHM's have been already obtained we think that the new lenses 
might be extremely important for 3D imaging.

In the rest of this paper we calculate the positions of the multiple
foci of the new lenses and the distribution 
of intensities among these foci.
Since the angles of incidence, reflection and refraction are equal to each 
other, a simple geometric construction gives the 
following equations for the positions
of the foci
\begin{align*}
z= a-m_0d\ {\rm and}\ z = \pm (2 d m -a),\ & m=m_0+1,m_0+2 ..., & m_0 -{\rm even (or\ zero)}\\
z = \pm (2 d m -a),\ & m=m_0,m_0+1 ...,& m_0 -{\rm odd},
\end{align*} 
where $m_0 = {\rm Int}[a/d]$, $a$ is the distance from the point source to the slab.

To study the intensities in the foci one should know 
the reflection and transmission coefficients at the RHM-LHM and LHM-RHM interfaces.
As has been shown by Veselago\cite{ve}
one can get them from the regular Fresnel expressions by substituting 
the absolute values of $\epsilon'$ and $\mu'$.
Then the reflection and transmission coefficients 
at the RHM-LHM interface $r$, $t$ and LHM-RHM interface $r'$, $t'$
have a form
\begin{equation}
\label{rt}
t = \frac{2 \epsilon}{\epsilon + |\epsilon'|},\ \ r=\frac{|\epsilon'|-\epsilon}{|\epsilon'| + \epsilon}
\end{equation}
\begin{equation}
\label{rt1}
t' = \frac{2 |\epsilon'|}{\epsilon + |\epsilon'|},\ \ r'=\frac{\epsilon -|\epsilon'|}{\epsilon + |\epsilon'|}
\end{equation}
These equations are valid for an arbitrary angle of incidence and polarization\cite{jac}.
The multiple scattering approach 
gives the following expression for the intensities in the foci in the case
when $a<d$
\begin{gather}
I(a)=(1-r^2) I_0\label{int_0_0}\\ 
I(2dm-a)= (1-r^2)^2 r^{4m-4}I_0 {\rm ,}\  m=1,2...\label{int_0_1}\\
I(-[2dm-a])= (1-r^2)^2 r^{4m-2}I_0{\rm ,}\  m=1,2...{\rm\ ,} \label{int_0_2}
\end{gather}
where $I_0$ is the intensity of the source.
The sum of the intensities over all foci at the right of the slab is 
$I_R = I_0 (1-r^2)/(1+r^2)$. The net intensity in the foci at the left of the slab is
$I_L = I_0 r^2 (1-r^2)/(1+r^2)$.
One can check that $I_R + I_L + I_{\rm lost} = I_0$, where $I_{\rm lost} = r^2 I_0$ 
is the 
intensity of the light which did not contribute to any of the foci and which is
the light reflected from the first interface.
Another energy conservation statement is that the intensity in the focus inside the 
slab $z=a$ is equal to the sum of the intensities over all foci at the left 
and at the right of the slab: $I_L + I_R = I(a) = (1-r^2) I_0$. 
At a given value of $n'$ lenses of the new type differ from each other 
by reflection coefficient $r$. For a specified refractive index $n'$ 
one might achieve 
different results by choosing proper values of $r$. To 
obtain maximum intensity in the closest to the lens focus
one should make $r$ close to $0$ 
(the limit $r=0$ corresponds to the 
VL with one focus only outside of the slab).
If $r$ close to $1$, the 
intensities change slowly from one focus to another, thus
one obtains an array of images with almost equal intensity.

Now let us study the three dimensional images produced by our lens and 
their spatial orientation relatively to the source (Fig. \ref{fig2}(A)).
One can see that the images at the left side of the slab are inverted, 
however the images at the right side have the same orientation as the source.
The image inside the slab is also inverted.

When the distance from the source to the slab satisfies the relation
$d<a<2d$,  
%the coordinates of the foci created by our lens (Fig. \ref{fig3}) are given by 
%\begin{equation}
%z = \pm(2 d m -a),\ m=1,2 ...
%\end{equation} and 
the intensities in the foci are 
\begin{gather}
I(2d-a) = (1-r^2) r^2 I_0\label{int_1_0}\\ 
I(2dm-a) = (1-r^2)^2 r^{4m-4}I_0 {\rm ,}\  m=2,3... \label{int_1_1}\\
I(-[2dm-a]) = (1-r^2)^2 r^{4m-2} I_0{\rm\ ,}\  m=1,2...\label{int_1_2}
\end{gather}
The sum of the intensities over all foci at the right of the slab is
$I_R = I_0 r^4 (1-r^2)/(1+r^2)$. The net intensity in the foci at the left of the slab 
$I_L = I_0 r^2 (1-r^2)/(1+r^2)$.
It is easy to show that $I_R + I_L + I_{\rm lost} = I_0$, where in this case 
$I_{\rm lost} = [r^2 + (1-r^2)^2]I_0$ is the 
intensity of the light which did not contribute to any of the foci.
As well as in the previously considered case of $a<d$ we have that
$I_L + I_R$ is equal to the intensity
in the focus inside the slab $I(2d-a) = r^2 (1-r^2)$.
At $r$ close to $1$ the intensities in the foci outside the slab change 
slowly from one focus to another.
Figure \ref{fig2}(B) also shows the 3D images and their orientations.
In this case the images at the left side of the slab are inverted, while the images at
the right side and the image inside the slab are not.
Note, that at $a>d$ the VL does not have any real images.

Thus, we proposed a new type of lenses on the basis
of the LHM. They are easier for manufacturing than the VL,
they have multiple foci, and they can produce 
3D images at the distances greater than the width of the LHM slab.

The work has been funded by the NSF grant DMR-0102964.

\bibliographystyle{apsrev}
\bibliography{index}


\end{document}